\documentstyle[aps,epsf,twocolumn]{revtex}
\input epsf
\begin{document}
\draft
\preprint{NSF-ITP-97}
\flushbottom
\twocolumn[
\hsize\textwidth\columnwidth\hsize\csname @twocolumnfalse\endcsname

\title
{Theory of Spin Fluctuations in Striped Phases of
Doped Antiferromagnetic Cuprates}
\author{Daniel Hone$^{1}$ and A.H. Castro Neto$^{2}$}
\bigskip
\address{$^{1}$
Institute for Theoretical Physics,
University of California,
Santa Barbara, California, 93106}

\address{$^{2}$
Department of Physics,
University of California,
Riverside, California, 92521 }
\date{\today}

\maketitle
\tightenlines
\widetext
\advance\leftskip by 57pt
\advance\rightskip by 57pt

\begin{abstract}
We study the properties of generalized striped phases 
of doped cuprate planar quantum antiferromagnets.
We invoke an effective, spatially
anisotropic, non-linear sigma model in two space dimensions.  
Our theoretical
predictions are in {\it quantitative} agreement with recent experiments
in La${}_{2-x}$Sr${}_x$CuO${}_4$ with
$0 \leq x \leq 0.018$.
We focus on (i) the magnetic correlation length, (ii) the staggered
magnetization at $T=0$ and (iii) the N\'eel
temperature, as functions of doping, using parameters determined 
previously and independently for this system.
These results support the proposal
that the low doping (antiferromagnetic) phase of the cuprates has a
striped configuration.
\end{abstract}

\vskip 0.5cm

]
\narrowtext
\tightenlines

There is no direct evidence for periodically structured striped phases in
doped antiferromagnetically ordered cuprates, such as
La$_{2-x}$Sr$_x$CuO$_4$ (with $ x < 0.02$).  Indeed,
recent neutron scattering measurements \cite{yamada}
show that the incommensurate magnetic peaks which are characteristic
of dynamic
stripe formation at higher hole concentrations $x$, seem to disappear
below $x$ of order 0.05, well above the antiferromagnetic regime.
Yet the growing theoretical literature\cite{outros} supporting the
tendency toward microscopic phase separation of holes in these strongly
correlated quasi-two-dimensional materials, as well as the experimental
observation\cite{experimental,stripe} of stripe phenomena in many
related systems, raises the
strong possibility that the doped antiferromagnetic cuprates will also be
characterized by such stripes.
Moreover, there is {\it indirect} evidence of linear,
or striped, features in the magnetic structure of these materials, including
the finite size scaling properties\cite{cho} of the N\'eel
temperature and uniform
magnetic susceptibility, and the successful interpretation\cite{borsa,prl}
of muon spin resonance ($\mu$SR) and nuclear quadrupole resonance (NQR)
experiments within models that presume a striped structure.

Those latter
theories are nominally based on a static periodic array of stripes, which
separate antiferromagnetic slabs, or ladders, at most weakly coupled to
one another across the stripes.  Since such regular arrays are not
observed in the neutron scattering experiments, we want to consider a
broader class of striped structures. These may include arrays with varying
separation between neighboring stripes, as suggested\cite{tranq2} by
the neutron scattering lineshapes in the related doped nickelates, or
dynamic behavior of the stripes, including amplitude fluctuations,
rigid translations, or the meanderings proposed\cite{zaanen} by Zaanen and
coworkers.  We also note that a magnetic phase domain, rather than anti-phase
boundary\cite{portugal}, which locally only suppresses, rather than
reverses the antiferromagnetic order parameter, would give at best only
a weak incommensurate scattering, even for a periodic array. It is then
our purpose here to develop a reliable effective field theory
which will predict the experimentally observable behavior while making
a minimum of assumptions about the details of the underlying structure,
beyond the demand that it corresponds on average to periodic
one-dimensional weakening of the antiferromagnetic exchange between
spins.

Encouraged and guided by the remarkable success of Chakravarty,
Halperin and Nelson\cite{chn} in describing {\it undoped} La$_2$CuO$_4$ by
just such an approach, we introduce a suitably anisotropic non-linear
sigma model parametrized so
as to reproduce the long wavelength behavior of the doped material in
the antiferromagnetically ordered phase.  Whatever the origin of the
average periodic modulation of exchange, the corresponding long
wavelength magnetic excitations from the antiferromagnetic ground
state, which determine the thermodynamics, will
have the same
character as those of a static array of stripes with the same period.
They are gapless (Goldstone modes, from spin rotational symmetry), with
a dispersion relation for the frequency squared which is analytic in
wave vector, giving a linear dispersion with anisotropic ``spin wave"
velocity, where the principal axes of the velocity tensor must be
parallel and perpendicular to the stripes (which lie along a crystal
axis\cite{stripe}).  Such behavior is described
in the continuum limit by the action of an effective anisotropic
non-linear sigma model,
\begin{eqnarray}
S_{eff} &=& \frac{1}{2}
\int_0^{\beta \hbar} d \tau \int dx \int dy
\left\{ S^2 \left[J_y \left(\partial_y \hat{n}\right)^2
+ J_x \left(\partial_x \hat{n}\right)^2 \right] \right.
\nonumber
\\
&+& \left.
\frac{\hbar^2}{2 a^2 (J_x+J_y)} \left(\partial_{\tau} \hat{n}\right)^2
\right\},
\label{effa1}
\end{eqnarray}
where $\hat n$ is a unit vector field. The symbols have been chosen to
suggest the continuum limit of an underlying effective {\it integer spin}
\cite{gap} Heisenberg hamiltonian on a square lattice with
lattice constant $a$, and with nearest neighbor exchanges $J_x$ and $J_y$.
In terms of these parameters the principal spin wave velocities are
\begin{eqnarray}
c_y^2 = 2 S^2 a^2 J_y (J_x+J_y)
\nonumber
\\
c_x^2 = 2 S^2 a^2 J_x (J_x+J_y).
\label{velocities}
\end{eqnarray}
The fundamental underlying anisotropy parameter is then equivalently the
ratio of the two exchange constants or of the two velocities,
\begin{equation}
\alpha = J_x / J_y.
\label{anisotropy}
\end{equation}
Its value characterizes the theory, but just how it decreases with
increasing doping concentration $x$ will have to be set later.

We note that in restricting the subsequent analysis to
these longest wavelength
excitations we ignore both the non-linear dispersion of these
acoustic modes and all optical branches associated with the
(average) superlattice in the $x$-direction.  As an estimate of the
lowest lying optical mode energies we can take the location of the
lowest gap introduced by a stripe superlattice in the hole-free
antiferromagnetic host, which occurs approximately at 
$J\pi\sqrt{2}/N_0$ for a superlattice
of period $N_0$ lattice constants in the host with exchange constant
$J$. Coulomb energy costs would seem to prevent values of $N_0$ much
greater than  20 to 30, giving  
a lowest optical mode energy greater than about 250 - 350 K. These 
energies lie above the temperatures of interest, and the neglect
seems reasonable.

From this point on we make use of well established techniques to
analyze the behavior of the field theory described by the action
(\ref{effa1}) so as to make predictions for the physical properties of
the system of interest.  First it is useful to rewrite (\ref{effa1})
more symmetrically by a dimensionless
rescaling of the variables:
$x'= (\alpha)^{-1/4} x \Lambda$,
$y'= (\alpha)^{1/4} y \Lambda$ ($\Lambda\sim 1/a$ is a momentum cut-off), and
$\tau'= \sqrt{2 (J_x+J_y) \sqrt{J_x J_y}} S a  \tau/\hbar$.
Then the effective action (\ref{effa1}) becomes
\begin{equation}
S_{eff} = {\hbar\over (2 g_0)} \int_0^{\hbar \Lambda \beta c_0}
d \tau' \int dx' \int dy'
\left(\partial_{\mu} \hat{n}\right)^2,
\end{equation}
where $\mu$ takes the values $ x',y',\tau'$,
\begin{equation}
g_0(\alpha) = \hbar c_0 \Lambda/\rho^0_s =
\left[2 (1+\alpha)/\sqrt{\alpha}\right]^{1/2}(a \Lambda)/S
\label{go}
\end{equation}
is~the~bare~coupling~constant, $c_0= [2 (J_x+J_y) \sqrt{J_x J_y}]^{1/2}
(a S)/\hbar$
the spin wave velocity and
$\rho^0_s = \sqrt{J_x J_y} S^2$ the classical spin stiffness of the rescaled
model. The original anisotropy is now hidden in the limits.
We started with a problem with a
finite bandwidth, a lower bound on length which requires us to
impose a cutoff in
the original continuum formulation.  The change of variables introduces
an anisotropy in the cutoffs.

The $\sigma$-model action, and the spin correlations it implies, can be
studied in the large $N$ limit ($N$ is the number of components of
$\hat n$), where a saddle point approximation becomes exact
\cite{largeN,assa}.
In the antiferromagnetically disordered phase the staggered spin-spin
static correlation function is given\cite{big} by
\begin{equation}
\langle\sigma(x,y)\sigma(0,0)\rangle = \frac{g_0(\alpha)}{r_s}e^{-mr_s},
\label{corrf}
\end{equation}
where $\sigma$ is the component of $\hat n$ in the ordering
direction, $r_s$ is the scaled length, $r_s^2 = x^2/\sqrt{\alpha}+
\sqrt{\alpha}y^2$, and $m$ is the inverse correlation length, given
formally by imposing the condition that the magnitude of the field 
$\hat n$ is unity
at each point.  At zero temperature this condition becomes
\begin{equation}
1 = 2g_0(\alpha)\int_0^{\alpha^{-1/4}}\frac{dk_y}{2\pi}\int_0^{\alpha^{1/4}}
\frac{dk_x}{2\pi}\frac{1}{\sqrt{k^2+m^2}}.
\label{selfc}
\end{equation}
We find that $m\ll \alpha^{1/4}$ over the full range of parameters
of interest,
in which case (\ref{selfc}) gives
\begin{equation}
m(\alpha) = 8\pi\left(\sqrt{\alpha}\over{1+\alpha}\right)^{1/2}
\left[\frac{1}{g_c(\alpha)} - \frac{1}{g_0(1)}\right],
\label{malpha}
\end{equation}
where
\begin{eqnarray}
g_c(\alpha) &=&
 \sqrt{8} \pi^{2} \sqrt{\alpha/(1+\alpha)}  \left\{
 \ln\left(\sqrt{\alpha}+\sqrt{1+\alpha}\right)\right.
\nonumber
\\
&+& \left. \sqrt{\alpha}\ln[(1+\sqrt{1+\alpha})/\sqrt{\alpha}]
\right\}^{-1},
\label{gca}
\end{eqnarray}
is the critical coupling constant of the theory: the ground state
is disordered\cite{gcnote} if $m(\alpha)>0$, or
$g_0(1)<g_c(\alpha)$.
It decreases monotonically with increasing anisotropy $1/\alpha$,
as the hole doping concentration $x$ grows.
Thus the system remains disordered at $T=0$ for $\alpha < \alpha_c$,  
where $\alpha_c$, the critical anisotropy, is defined by the
condition $g_c(\alpha_c)=g_0(1)$.  We will see below, as this
suggests, that
as $\alpha$ approaches $\alpha_c$ from {\it above} with increasing doping,
this is also the value where the spin stiffness --- and the three
dimensional N\'eel temperature --- vanishes.  Numerically,
$\alpha_c\approx 0.047$.

For sufficiently weak anisotropy ($1 > \alpha > \alpha_c$) that the system
orders magnetically at zero temperature, we can take advantage of
the power of renormalization group techniques for systems with
diverging correlation lengths, to understand the corrections to the
classical limit above.  As always, we introduce an explicit length
scale, $a\rightarrow \Lambda a = e^{-\ell}$, where $a$ is the original
lattice parameter. Then the normalization condition  on the
field $\hat n$ at $T=0$ becomes
\begin{equation}
\bar\sigma^2 = 1 - \frac{2g_0(\alpha)}{\Lambda}\int_0^{\alpha^{-1/4}
\Lambda}\,\frac{dk_y}{2\pi}\int_0^{\alpha^{1/4}\Lambda}\,\frac{dk_x}{2\pi}\,
\frac{1}{k},
\label{sigbar}
\end{equation}
where $\bar\sigma$ is the average value of the order parameter (staggered
magnetization), and the second term on the right hand side represents
the spin flip fluctuations away from perfect N\'eel order.
The RG procedure can
be carried out in various ways, including (see, e.g., \cite{chn})
explicit integration over large momentum values and suitable re-scaling
to restore Eq. (\ref{sigbar}) to its original form.  Instead we
introduce\cite{fradkin} the explicit re-scaling function ${\cal Z}
(\Lambda)$:
$\bar\sigma = {\cal Z}(\Lambda)M$, and $g_0(\alpha) = {\cal Z}(\Lambda)g_R
(\alpha)$, which renders (\ref{sigbar}) renormalizable, with the
renormalization equation,
\begin{equation}
\frac{dg_R}{d\ell} \approx -g_R +\frac{g_0(1)}{g_0(\alpha)g_c(\alpha)}g_R^2,
\label{beta}
\end{equation}
which can be integrated from the bare value $g_0(\alpha)$ at $\ell=0$
to the fully renormalized value at $\ell\rightarrow\infty$.  The stable
fixed point is at $g_R = 0$, justifying keeping only the lowest two
powers of $g_R$ in the equation (\ref{beta}), a one-loop approximation.
The result then gives
for the renormalized spin stiffness,
\begin{equation}
\rho_s(\alpha) \rightarrow \left[ \frac{\hbar c}{ae^\ell g_R(\ell,\alpha)}
\right]_{\ell\rightarrow\infty} = \rho_s^0(\alpha)\left[
1 - \frac{g_0(1)}{g_c(\alpha)}
\right],
\label{stiff}
\end{equation}
reduced from its classical value $\rho_s^0(\alpha)$ for fixed anisotropy
$\alpha$ in such a way that it vanishes, as foreseen above,
at the critical value $\alpha=\alpha_c$.

Now we turn to prediction of and comparison with experimental quantities,
including correlation length $\xi(\alpha,T)$, zero point
magnetization $M_s(\alpha)$, and N\'eel temperature $T_N(\alpha)$.
We know from studies of the undoped system that while the
one-loop calculation accurately predicts the leading exponential
dependence on inverse temperature of $\xi(T)$, its results
are not very good for the prefactor of that exponential or the algebraic
temperature dependent corrections, and the same will surely be true
for the doped system, with $\alpha < 1$.  Therefore we use an interpolation
formula between the exact result of Hasenfratz and Niedermayer\cite{exact}
for the nonlinear sigma model in the neighborhood of magnetic order, and
the result for the quantum critical regime, where $\xi\propto T^{-1}$:
\begin{equation}
\xi(T,\alpha) \approx \left(\frac{e\hbar c_0}{4}\right)
\frac{e^{2 \pi \rho_s(\alpha)/k_B T}}{4\pi\rho_s(\alpha)+ k_B T}.
\label{interpol}
\end{equation}
This gives
excellent agreement \cite{chn} with experiment in the pure case, $x=0$.
The experimental data\cite{keimer} for $\xi^{-1}(x,T)$ have been
interpreted phenomenologically, not according to (\ref{interpol}), but
as the sum of the pure system function plus a constant depending only
on doping concentration $x$: $\xi^{-1}(x,T) = \xi^{-1}(0,T) +
\xi^{-1}(x,0)$. But only the pure ($x=0$) experimental results
are reliably in the antiferromagnetic region we treat here.
The other curves, with nominal values of $x =$ 0.02,
0.03 and 0.04, correspond most likely to the cluster spin glass phase.
Of course, to make a direct comparison with the experiments, yet to be
done, in the region $x<0.02$, we need the connection between the
anisotropy parameter $\alpha$ and the hole concentration $x$.  We can
do this indirectly, by comparing experimental properties, for example,
as a function of N\'eel temperature, as we will do below for the
zero point staggered magnetization.  One additional parameter is still
required, though, for comparison with (\ref{interpol}).  The spin
stiffness (\ref{stiff}) depends through $\rho_s^0$ on the product
$J_xJ_y$, as well as the ratio $\alpha$.  If we take that product, for
example, to be independent of $x$, or $\alpha$, then the curves for
$\xi^{-1}(T)$ for doped and undoped systems will be found to cross, as
seems to be suggested by the experiments \cite{keimer} for
$x =$ 0 and 0.02.

In general, the staggered magnetization $M_s$ depends on the short,
as well
as the long wavelength physics of the problem, so we can't use the
results at $T=0$ of the long distance nonlinear sigma model theory
directly. Instead, as in \cite{chn}, we use the asymptotic long
distance behavior of the equal time spin-spin correlation function:
$\langle\sigma(x,y)\sigma(0,0)\rangle \rightarrow (M_s/M_0)^2$ as
$x,y \rightarrow\infty$, where $M_0$ represents perfect N\'eel spin
alignment.  To do this we evaluate (\ref{corrf}) at the point where
the scaled length $r_s$ is equal to the Josephson correlation length
$\xi_J = \hbar c/\rho_s$, which separates long from short
wavelength scales. On the one hand, this is large enough for the
correlation function to have reached its desired asymptotic
behavior, but it is still short enough that the (exponential) decay
due to the long wavelength fluctuations at $T=0^+$ where (\ref{corrf})
holds has not yet become effective.  Thus we find,
\begin{equation}
\frac{M_s(\alpha)}{M_s(1)} = \sqrt{\frac{1-g_0(1)/g_c(\alpha)}{
1-g_0(1)/g_c(1)}}.
\label{ms}
\end{equation}

As usual\cite{imry} for quasi-two-dimensional systems, we make
use of the long range order parameter correlations that have developed
in the plane above the ordering temperature $T_N$, to call on the
validity of mean field theory for the weak exchange $J_{\perp}$
in the third
dimension in establishing 3-d magnetic order at a finite temperature.
A simple physical interpretation of the result is that $T_N$ is the
temperature at which the energy of thermal fluctuations, $k_BT_N$, becomes
sufficient to flip the spins in a region of linear dimension of the
order of the 2-d correlation length, $\xi(T_N)$. Since the number of
spins in this region is proportional to $(\xi/a)^2$,
and the relative staggered magnetization in the
region is given by $M_s/M_0$, we estimate
\begin{equation}
k_B T_N(\alpha) \approx
J_{\perp} \left(\frac{\xi(T_N,\alpha)}{a} \frac{M_s}{M_0}\right)^2.
\label{heur}
\end{equation}
This expression has been used previously~\cite{chn} to
estimate $J_\perp/k_B\approx 0.01$ K from the experimental $T_N$ of the pure
material. Since $M_s/M_0 < 1$, this gives $(\xi/a) > 10$ for $T_N > 1$ K,
suggesting that this mean field theory is reasonable for $T_N$ greater than
a few kelvin.

Though the predicted staggered magnetization (\ref{ms}) and the  N\'eel
temperature (\ref{heur}) both depend on the anisotropy parameter $\alpha$,
whose explicit dependence on doping $x$ is not established by the model,
we can eliminate this dependence between the two relations, plotting
$M_s$ {\it versus} $T_N$, each normalized to the corresponding undoped
($x=0$) value, as is done in Fig. 1.  The experimental values are those
of Ref. \cite{borsa}.  We note, in particular, that the zero point
magnetizations are extrapolated values.  As is explained in \cite{borsa},
there is a change of behavior around $T = $ 30K for all values of doping.
The authors have interpreted this as some sort of freezing of the holes,
which may be binding to the donor impurities, establishing static
charge density waves within the stripes, or some other change in
behavior.  In any
case, the reported values of $M_s$ are those extrapolated from the
observed magnetization curves at temperatures above this ``freezing".
We emphasize that the comparison in Fig. 1 has {\it no} adjustable
theoretical parameters, and the agreement with experiment is
excellent.  This, and the further comparisons made in \cite{prl}, 
then supports a generalized stripe picture of the antiferromagnetic 
doped cuprates.

\begin{figure}
\hspace{0cm}
\epsfxsize=7cm\epsfbox{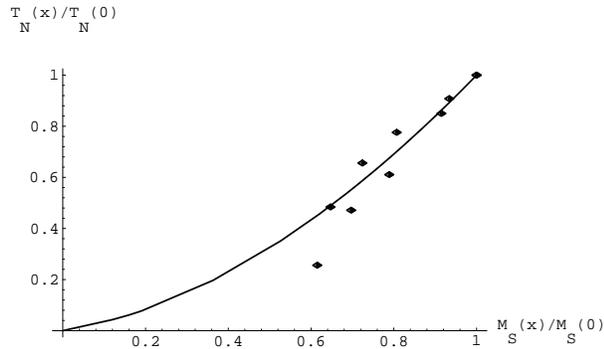}
\caption{Neel temperature versus staggered magnetization normalized to
the undoped values. Line: theory; diamonds: experiment.}
\label{imp}
\end{figure}

We are indebted to S.A. Kivelson for introducing us to this
problem and for his illuminating comments and suggestions throughout
the work. We thank A. Balatsky, S. Chakravarty, C. Di Castro, and
C. Castellani for valuable coments, and
F. Borsa for discussion of his and other experiments.
We also acknowledge support by NSF Grant PHY 94-07194.


\begin{references}
\bibitem{gcnote} Note that the condition is expressed relative
to the fixed undoped coupling constant $g_0(1)$, given by Oguchi's
renormalized spin wave theory\cite{oguchi,chn} as $g_0(1)\approx 9.54$.
By using Eq. (\ref{go}), which gives $g_0(\alpha)/g_0(1)$ we could
readily write this directly as an inequality for the actual coupling constant
of the anisotropic system for each value of $\alpha$, a
more obvious condition physically, but it is
algebraically simpler to do it in terms of $g_0(1)$.
\end{references}
\end{document}